\documentclass{aa}  

\bibpunct{(}{)}{;}{a}{}{,}
\usepackage{graphicx}
\usepackage{txfonts}
\usepackage{subfigure}
\usepackage{xcolor}
\usepackage{float}
\usepackage{natbib}
\usepackage{amsmath}

\DeclareRobustCommand{\VAN}[3]{#2}
\let\VANthebibliography\thebibliography
\def\thebibliography{\DeclareRobustCommand{\VAN}[3]{##3}\VANthebibliography}

\begin{document}

    \title{VLBI Imaging of high-redshift galaxies and protoclusters at low radio frequencies with the International LOFAR Telescope}
    \titlerunning{VLBI Imaging of high-redshift galaxies and protoclusters at low radio frequencies with the ILT}

   \author{C. M. Cordun\inst{1}\fnmsep\thanks{Corresponding author \email{cordun@mail.strw.leidenuniv.nl} }
          \and
          R. Timmerman\inst{1}
          \and
          G. K. Miley\inst{1}
          \and
          R. J. van~Weeren\inst{1}
          \and
          F. Sweijen \inst{1,2}
          \and
          L. K. Morabito\inst{3,4}
          \and 
          H. J. A. R\"ottgering\inst{1}
}

   \institute{Leiden Observatory, Leiden University, P.O. Box 9513, 2300 RA Leiden, The Netherlands
             \and
            ASTRON, Netherlands Institute for Radio Astronomy, Oude Hoogeveensedijk 4, 7991 PD Dwingeloo, The Netherlands
             \and
             Centre for Extragalactic Astronomy, Department of Physics, Durham University, Durham DH1 3LE, UK
             \and
            Institute for Computational Cosmology, Department of Physics, University of Durham, South Road, Durham DH1 3LE, UK\\
             }

   \date{Received XXX; accepted YYY}

 
  \abstract
   {It has long been known that luminous, ultra-steep spectrum radio sources are preferentially associated with massive galaxies at high redshifts. Here we describe a pilot project directed at such objects, to demonstrate the feasibility and importance of using LOFAR to study the most distant forming massive galaxies and protoclusters. We have successfully imaged four high-redshift ($z>2$) high-luminosity radio galaxies with sub-arcsecond resolution, at 144~MHz, using the International LOFAR Telescope (ILT). Our targets were 4C\,41.17 ($z=3.8$),  the \textquotedblleft Anthill\textquotedblright, B2\,0902+34 ($z=3.4$), 4C\,34.34 ($z=2.4$) and 4C\,43.15 ($z=2.5$). Their low-frequency morphologies and the spatial distributions of their low-frequency spectral indices have been mapped, and compared with available optical, infrared, and X-ray images. Both for the Anthill at $z$ = 3.8 and B2\,0902+34 at $z$=3.4, the location of the steepest radio emission coincides with the Ly$\alpha$--emitting ionized gas halo.   
   Our pilot project demonstrates that, because of its outstanding sensitivity and high angular resolution at low frequencies, the ILT is a unique facility for studying the co-evolution and interaction of massive galaxies, galaxy clusters, and supermassive black holes in the early Universe.}


   \maketitle
%

\section{Introduction}

High-redshift radio galaxies (HzRG) are among the largest, most luminous and most massive galaxies in the early Universe \citep{chambers1988high,miley1989high, mccarthy1993high,blundell1999inevitable, miley2008distant, miley2009high, saxena2019nature}. They are likely progenitors of massive elliptical/cD galaxies, located at the centers of low-redshift galaxy clusters \citep{fall1977survival}. HzRGs are therefore unique laboratories for studying the formation of massive galaxies, protoclusters, and active galactic nuclei (AGN) and their interaction with the surrounding environment.

Recently it has been suggested that the emergence of massive galaxies, protoclusters, and supermassive black holes is interconnected \citep{ricarte2019clustering}. The observational evidence for the co-evolution of these three inhabitants of the early Universe includes the alignments that have been observed between their radio jets and the associated optical continuum, line-emitting ionized gas, molecular gas, and X-ray emission \citep{miley2008distant}. In several cases, the alignments indicate that star formation in the protocluster is triggered by the interaction of the radio-emitting jet with the surrounding intergalactic medium \citep{1989ApJ...342L..59D, Bicknell_2000, nesvadba2006extreme}. Well-known cases of such alignments are the spectacular \textquotedblleft Spiderweb\textquotedblright  radio protocluster at \(z=2.2\) (\citet{miley2008distant}) and the 4C\,41.17 radio galaxy at \(z=3.8\). Because of the similarity of 4C\,41.17 with the Spiderweb, we have named it the \textquotedblleft Anthill\textquotedblright galaxy cluster. Recently, an overdensity of X-ray AGN was detected in the field of the Spiderweb that is an order of magnitude greater than that of the field \citep{tozzi2022700}, implying  that this forming protocluster is populated with fast-growing super-massive black holes (SMBHs) and that the early evolution of protoclusters, their central massive galaxies and the associated super-massive black holes (SMBHs) are closely connected.

Although distant protoclusters have been detected by other methods \citep[e.g.,][]{da2021identifying, casey2015massive, adam2015pressure, toshikawa2018goldrush, uchiyama2022wide}, surveys at low radio frequencies is still one of the most powerful tools for finding them \citep{wylezalek2013galaxy} and studying the role of AGNs and SMBHs in their formation and evolution.

A specific characteristic of HzRGs that enables them to be more easily identified in the radio band is the correlation between their spectral index ($\alpha$) and redshift ($z$) - the $\alpha\sim z$ relation \citep{tielens1979westerbork}. This empirical law indicates that the steeper the overall spectrum is (from low, MHz frequencies to high, GHz frequencies), the higher its redshift is likely to be \citep{chambers19904c}. Hence ultra-steep spectrum (USS) radio sources  pinpoint the most massive galaxies at the largest redshifts. The origin of the $\alpha \sim z$ relation is still not fully understood. 
One possibility is that the steeper spectral index is due to increased confinement of the radio lobes by the denser protocluster environment at higher redshifts. This could cause the radio surface brightness of the lobes to remain relatively high, resulting in increased spectral steepening due to energy losses \citep{klamer2006search}.
A second explanation is that it originates from a K-correction effect coupled with the concave radio spectra of HzRGs. It has been suggested that about 50\% of the $\alpha \sim z$ relation gradient can be explained by such an effect \citep{ker2012new}.
A third explanation is that the steeper radio spectrum at higher redshift may be related to increased inverse Compton losses due to up-scattering of the cosmic microwave background (CMB) photons by the radio synchrotron emitting relativistic electrons. Note that the CMB energy density increases as $(1+z){^4}$ \citep[e.g.,][]{carilli2022x}. 
A study by \citet{morabito2018investigating} found that inverse Compton losses in flux density limited surveys are enough to reproduce the observed $\alpha\sim z$ relation. Evidence that inverse Compton X-ray radiation of the surrounding environment is directly related to the steep radio spectra has also been found in multiple studies \citep[e.g.,][]{jimenez2021extended}. 

Further investigation of the poorly understood processes that play a role in building massive galaxies, protoclusters, and supermassive black holes (SMBHs) are fundamentally important.  In particular, to investigate the interaction of radio jets produced by the nuclear SMBHs of massive central galaxies with other protocluster constituents, making high resolution maps of their radio intensities and spectra at low frequencies are essential - the topic of this paper. 

The International Low Frequency Telescope \citep[ILT;][]{vanhaarlem13}, with its pan-European baselines is capable of achieving sub-arcsecond resolution in its HBA mode (120 - 168~MHz). This is an unprecedentedly high spatial resolution at such low frequencies. Because HzRGs (i) usually have steep spectra ($\alpha<-0.9$) and (ii) have an angular size distribution peaked at $\sim$ 7\arcsec, with 80\% of known HzRGs having angular sizes between 5\arcsec and  20\arcsec \citep{carilli1997radio}, the ILT is uniquely equipped for detecting and mapping HzRGs.  

Making images at the full resolution of LOFAR is challenging. The ILT is an interferometric array, with $\sim$ 40,000 small antennas concentrated in 52 stations, 38 of them in the Netherlands, 6 in Germany, 3 in Poland, and single stations in France, Great Britain, Ireland, Latvia and Sweden.  The total effective collecting area is $\sim$ 300,000 m$^2$ and the ILT baselines extend beyond $\sim$ 2000 km. The stations are connected by fiber optical cables. In its HBA mode, their digitized signals are distributed over 16 frequency channels in each of 240 bands = 3840 frequency channels between 120 and 168~MHz and combined via a supercomputer at the University of Groningen.
 The continuously changing ionospheric perturbations at low frequencies and the large number of ILT baselines and frequency channels result in huge datasets that need to be calibrated and analyzed \citep{morabito2022sub}. 
 
 Here we describe a pilot project to investigate the feasibility of carrying out sub-arcsecond radio imaging at 144~MHz with the ILT of a small sample of HzRGs, using the images to make high-resolution radio spectral distribution studies and comparing these with maps at non-radio wavelengths. The eventual goal is to investigate the interaction of radio jets with the other constituents of forming galaxies and protoclusters and its importance to galaxy and cluster evolution. Our pilot study is based on observations of three targets covered by the LOFAR Two-meter Sky Survey \citep[LoTSS;][]{shimwell2019lofar}, supplemented by a fourth HzRG target, 4C\,43.15 at \(z=2.5\), that has already been mapped at sub-arcsecond resolution using the ILT by \citet{sweijen2022high}. Our intention is to eventually extend this pilot study to a larger sample.

Throughout this paper, we assume a  $\rm \Lambda$CDM cosmology, with the following cosmological parameters: $ \rm H_0 = 70 \  km \ s^{-1} \ Mpc^{-1}$, $\Omega_{ {m}} = 0.3$  and  $\rm \Omega_\Lambda = 0.7$, where $\rm H_0$ is the Hubble parameter, $\Omega_{ {m}}$, the matter density parameter, and $\rm \Omega_\Lambda$, the dark energy density parameter. Moreover, all the images are made using the J2000 coordinate system and we define the spectral index as S $ \rm \propto \nu^\alpha$, where $\rm S$ is the flux density, and $\rm \nu$, the frequency. We shall refer to spectra that have $\alpha < -1$  \textquotedblleft ultra-steep\textquotedblright (USS).


\section{The HzRGs targets}

Our targets are listed in Table \ref{table:ILT_projects}. All of them were selected to have archival measurements at higher frequencies with the Very Large Array (VLA), allowing ILT-VLA spectral index distributions to be mapped.

\begin{itemize}
    \item {4C\,41.17  \textquotedblleft The Anthill\textquotedblright}: The 4C41.17 radio source is associated with a spectacular massive galaxy at redshift \(z=3.8\) \citep{miley2008distant}. It was discovered because of its high radio luminosity and ultra-steep spectral index by \citet{chambers19904c}. Deep imaging with the HST revealed that the host galaxy is a spectacularly large clumpy group of  faint galaxies that appear to be merging with the massive host galaxy. The HST observations provided important early evidence favoring hierarchical galaxy formation models.  \par
    \setlength{\parindent}{15pt}

    In several of its properties, the Anthill closely resembles the well-studied Spiderweb Galaxy/protocluster at \(z=2.2\) \citep{miley2008distant}. Both are among the most massive, clumpiest, and largest galaxies known, consistent with cD galaxies in the process of formation. Both are surrounded by giant Lya-emitting ionized halos whose sizes exceed 100 kpc. Furthermore, both the Spiderweb \citep{di2023forming} and the Anthill \citep{emonts2023cosmic} are connected with large gas streams, extending by > 100 kpc. Spectroscopically confirmed protocluster-sized galaxy overdensities have been detected around the Spiderweb and several other similar high-luminosity radio galaxies \citep[e.g.][]{venemans2007protoclusters, tozzi2022700}. In the case of the Anthill, no protocluster-sized galaxy overdensity has yet been confirmed \citep{ivison2000excess, greve2007wide,wylezalek2013galaxy} and there are no Chandra X-ray observations of comparable depth to those of the Spiderweb \citep{tozzi2022700}. The fact that the Anthill is a "twin" of the Spiderweb in so many ways, is evidence that it is also the brightest cluster galaxy in the process of formation. However, the associated protocluster galaxies have not yet been detected.
    
     \item {B2\,0902+34}: This HzRG has a redshift of \(z=3.4\), a relatively flat optical spectral energy distribution, and was suggested to be a protogalaxy undergoing its first episode of star formation \citep{lilly1988discovery,eales1993evidence,pentericci1999hst}. The radio/optical structure of this galaxy has been described as  \textquotedblleft bizarre\textquotedblright, with a flat-spectrum radio core located in a  \textquotedblleft valley\textquotedblright between 2 optical peaks \citet{carilli1995bizarre}. This HzRG is also exceptional in that it is one of the only \(z>2\) radio sources that has been observed to have an associated narrow HI absorption line \citep{1991PhRvL..67.3328U, chandra2004associated}.
     
     \item {4C\,34.34}: This HzRG has a redshift of \(z=2.4\). It has been mapped with the VLA at higher frequencies \citep{law1995vla} and is, therefore, a suitable target for low-frequency spectral index mapping with the ILT observation. However, besides its optical spectrum and a Spitzer observation which places it in a highly dense environment \citep{wylezalek2013galaxy}, there are no other optical, infrared, or X-ray data available for this source.
     
     \item {4C\,43.15}: High resolution LOFAR ILT maps of this HzRG have already been published, both with the HBA \citep{sweijen2022high} and the LBA \citep{morabito2016lofar}. It has a redshift of \(z=2.4\), a bright Ly\(\alpha\) halo \citep{villar2003kinematically, motohara2000infrared}, and it is located in a dense region, but with a lower overdensity than most HzRGs \citep{wylezalek2013galaxy}.
\end{itemize}

\section{Data reduction} \label{sec:data_reduction}

\subsection{Producing sub-arcsecond Images from the International LOFAR Telescope}

 Details of the ILT observations are summarized in Table \ref{table:ILT_projects}. Each target was observed for a total of 8 hours in a frequency band between 120~MHz and 168~MHz. We shall henceforth refer to these data by their central frequency as  \textquotedblleft144~MHz\textquotedblright \  data. In addition, a 10-minute long observation of the standard calibration source (3C\,295) was taken before and after every target observation.

\begin{table*}[hbt]
\centering
\caption{Targets selected for 144~MHz sub-arcsecond mapping with the ILT, with properties taken from \citet{miley2008distant}, ILT project details for each target, and properties of the resultant ILT images.}
\label{table:ILT_projects}
\begin{tabular}{lllll}
\hline
Object &
  4C 41.17 (\textquotedblleft Anthill\textquotedblright) &
  B2 0902+34 &
  4C 34.34 &
  4C 43.15 \\ \hline
RA (J2000)                                                   & 06h50m51.2s       & 09h05m30.1s       & 11h16m30.4s       & 07h35m22.5s       \\
DEC (J2000)                                                  & +41$^\circ$30'31" & +34$^\circ$07'57" & +34$^\circ$42'24" & +43$^\circ$44'25" \\
Redshift                                                     & 3.792             & 3.395             & 2.400             & 2.429             \\
Integrated spectral index $\rm \alpha _{144 \ MHz}^{1.4 \ GHz}$         & -1.30             & -0.93             & -0.93             & -1.10             \\
Angular size at 144 MHz (arcseconds) &
  14.3" &
  11.5" &
  17.3" &
  12.3" \\
Linear size (kpc) & 192               & 150               & 230               & 163               \\
Project code                                           & LC6\_021\tablefootmark{2}          & LT10\_010\tablefootmark{1}         & LT10\_010\tablefootmark{2}         & LT5\_006\tablefootmark{2}          \\
SAS Id                                                 & 558416            & 657262            & 746346            & 427100            \\
Number of available international stations & 50 & 48 & 50 & 47\\
ILT resolution (arcseconds) &
  0.2" $\times$ 0.2" &
  0.4" $\times$ 0.2" &
  0.3" $\times$ 0.2" &
  0.4" $\times$ 0.4" \\
RMS noise level ($\rm \mu Jy \ beam^{-1}$)&
  180 &
  134 &
  204 &
  170 \\ \hline
\end{tabular}

\tablefoot{
\tablefoottext{1}{LoTSS archive \citep{shimwell2017lofar}}
\tablefoottext{2}{Dedicated project additional to LoTSS}}

\end{table*}

Making images with the full resolution of the International LOFAR Telescopes is complex and computer-intensive. Unlike most other radio telescopes, because LOFAR is a dipole array with a wide field of view and includes a large number of baselines at low frequencies, the data volume is extremely large. ILT data reduction involves three main steps \citep{sweijen2022high, morabito2022sub}. First, the target and its calibrator need to be processed. During this stage, the phase effects due to clock drifts are separated from those due to the ionosphere \citep{2021selfcal, van2016lofar}. Secondly, the data must be processed through the ILT long baseline pipeline to calibrate the international stations. Thirdly, high-resolution images are produced using a self-calibration procedure.

The first stage of the reduction made use of the \textsc{Prefactor} software package \citep{van2016lofar,williams16,de2019systematic} to remove instrumental effects and minimize radio-frequency interference (RFI) by flagging the data using the \textsc{AOFlagger} package \citep{offringa13, offringa15}. More specifically, based on the geometry of the array and by comparing our data to a model of the calibration source, we derived corrections for the polarization alignment, the Faraday rotation, the sensitivity as a function of frequency within the bandpass and clock offsets between the stations. After applying these corrections, the remaining low-level RFI was flagged. Finally, phase corrections for the short-baseline (Dutch) stations were derived from a sky model obtained from the  Tata Institute of Fundamental Research (TIFR) - Giant Metrewave Radio Telescope (GMRT) Sky Survey \citep[TGSS,][]{intema2017gmrt}. This calibrated data was averaged to 8~seconds per integration with frequency channels of 98~kHz.

The second stage involved extending the calibration from the Dutch stations to the pan-European ILT, using the LOFAR-VLBI pipeline \citep{morabito2022sub}. The calibration solutions previously derived based on the calibrator source were also applied to the long-baseline ILT data. Then, for each observation, a bright unresolved source from the Long-Baseline Calibrator Survey \citep[LBCS,][]{jackson16, jackson2022sub} was used to derive fast phase and slow complex gain calibration solutions, so as to correct as much as possible for perturbations to the wavefronts primarily caused by the ionosphere. 

The third stage of the reduction focused on self calibration, using the procedure developed in \cite{2021selfcal}. In all three of the newly analysed sources, the target was known to possess sufficiently bright compact structure to act as an in-field calibrator based on the Long-Baseline Calibrator Survey \citep[LBCS,][]{jackson16, jackson2022sub}. Because the core stations had already been calibrated, these could be phased-up to narrow the effective field of view, thereby minimizing the distorting influence on the images from nearby radio sources. The starting model for the self-calibration was derived from the VLA image of the relevant target, (see Sect.~\ref{sec:vla}). To derive the optimum set of calibration solutions, we iterated with multiple cycles of self-calibration. In each cycle, we imaged and cleaned the target field using WSClean \citep{offringa-wsclean-2014,offringa-smirnov-2017} and used the deconvolved image as an updated model to derive improved calibration solutions, with the weighting parameters selected to give the best compromise between sensitivity and resolution \citep[e.g.,][]{briggs1995new}.

The final angular resolution achieved at 144~MHz was $\sim0.3$" (FWHM) for all four targets, the FWHM of the primary beam, $\sim2$~arcminutes, and the maximum recoverable scale is $\sim$50 arcseconds. An overview of the image properties and noise levels obtained is given in Table~\ref{table:ILT_projects}.

\subsection{Complementary VLA images and low-frequency spectral index maps}
\label{sec:vla}

\begin{table*}[ht]
\centering
  \caption{Complementary VLA observations used in this project.}
  \label{table:VLA} 
\begin{tabular}{cccccc}
\hline
\multicolumn{2}{c}{Object}             & 4C 41.17                         & B2 0902+34           & 4C 34.34 & 4C 41.15             \\ \hline
\multicolumn{2}{c}{Project code} &
  \multicolumn{1}{c}{\begin{tabular}[c]{@{}c@{}}AM0205, AC0234,\\  AC0316, AI0108, \\ AM 107, AM 0189\end{tabular}} &
  \multicolumn{1}{c}{\begin{tabular}[c]{@{}c@{}}AV0157, AB0511, \\ AC0316\end{tabular}} &
  \multicolumn{1}{c}{AV0157, AR0287} & \multicolumn{1}{c}{AK410}\\
\multicolumn{2}{c}{Observing time (h)} &                                  &                      &                      \\
                          & L band     & A config. - 8.4                  & A config. - 2.1      & A config. - 0.5  & -    \\
\multicolumn{1}{l}{} &
  C band & A config. - 2.1 &  A config. - 2.7 & A config. - 0.5 & -\\
  & &  B config. - 0.3 & B config. - 0.5 & B config. - 1.3 & - \\
  
\multicolumn{1}{l}{}      & U band     & B config. - 1.6 & -                    & -     & -               \\
 & & C config. - 0.6 & -  & - & - \\
\multicolumn{1}{l}{}      & X band     & -  & -                    & -    & A config - 0.7               \\
\multicolumn{2}{l}{Resolution}         & \multicolumn{1}{l}{}             & \multicolumn{1}{l}{} & \multicolumn{1}{l}{} \\
                          & L band     & 1.2" $\times$ 1.0"               & 1.1" $\times$ 1.0"   & 1.3" $\times$ 1.1"  & - \\
                          & C band     & 0.4" $\times$ 0.4"               & 0.5" $\times$ 0.4"   & 0.6" $\times$ 0.5"  & - \\
                          & U band     & 0.4" $\times$ 0.3"               & -                    & -   & -\\
                          & X band     & -               & -                    & -   & 1.05" $\times$ 0.83"
\end{tabular}
\end{table*}

To produce maps of spectral indices and starting models for the previously discussed self calibration stage of the ILT imaging, we used complementary data at higher frequencies from the Very Large Array (VLA) archive, as summarized in Table~\ref{table:VLA}. The array configurations were selected to optimally match the ILT angular resolutions. The VLA observations generally consisted of a single scan on a primary calibrator for flux scale and bandpass calibration and multiple short scans on a secondary calibrator between target observations for phase referencing.

The reduction and calibration of the VLA data was carried out using the Common Astronomy Software Application \citep[\textsc{CASA};][]{mcmullin07}. First, the \textsc{TFCrop} algorithm was used to detect and flag RFI. Calibration tables were then applied to correct for antenna position offsets and gain elevation curves. Next, the primary calibrator was used to set the absolute flux scale, and the secondary calibrator was applied to derive phase corrections for predominantly tropospheric perturbations. These were interpolated to the target scans and applied to the data together with the previously derived calibration tables.

Then, the target data were self-calibrated using WSClean \citep{offringa-wsclean-2014} for imaging and cleaning, and CASA for deriving updated calibration tables. The self-calibration process initially only derived updated phase calibration tables. When these converged, the self-calibration switched to deriving both phase and amplitude corrections. The PyBDSF software \citep{2015ascl.soft02007M} was applied to the final VLA images to derive starting models for the ILT self-calibration.

The final angular resolution achieved was $\sim$ 1.2" (FWHM) at 1.4~GHz (L band), $\sim$0.5" (FWHM) at 6~GHz (C band), $\sim$0.9" (FWHM) at 10~GHz (X band) and $\sim$0.4" (FWHM) at 15~GHz (U band), for all four targets.  An overview of the image properties is given in Table~\ref{table:VLA}.

The VLA and ILT images were convolved to similar angular resolutions using CASA and aligned with the \textsc{FITS\_tools} Python library. Then, low-frequency spectral index maps of each targeted source were made using these images. All pixels whose significance was smaller than $ 3 \sigma $ have been flagged.

\section{Results}
The results are here presented and discussed separately for each of our four targets. We show images of the high-resolution 144~MHz ILT emission, the 144 - 1420~MHz spectral index maps, and where available, superpositions of the radio data on optical and X-ray images.

\subsection{4C\,41.17  \textquotedblleft The Anthill\textquotedblright \  at \(z=3.8\)}

\begin{figure}[ht]

\begin{subfigure}
\centering
  \includegraphics[width=0.88\linewidth]{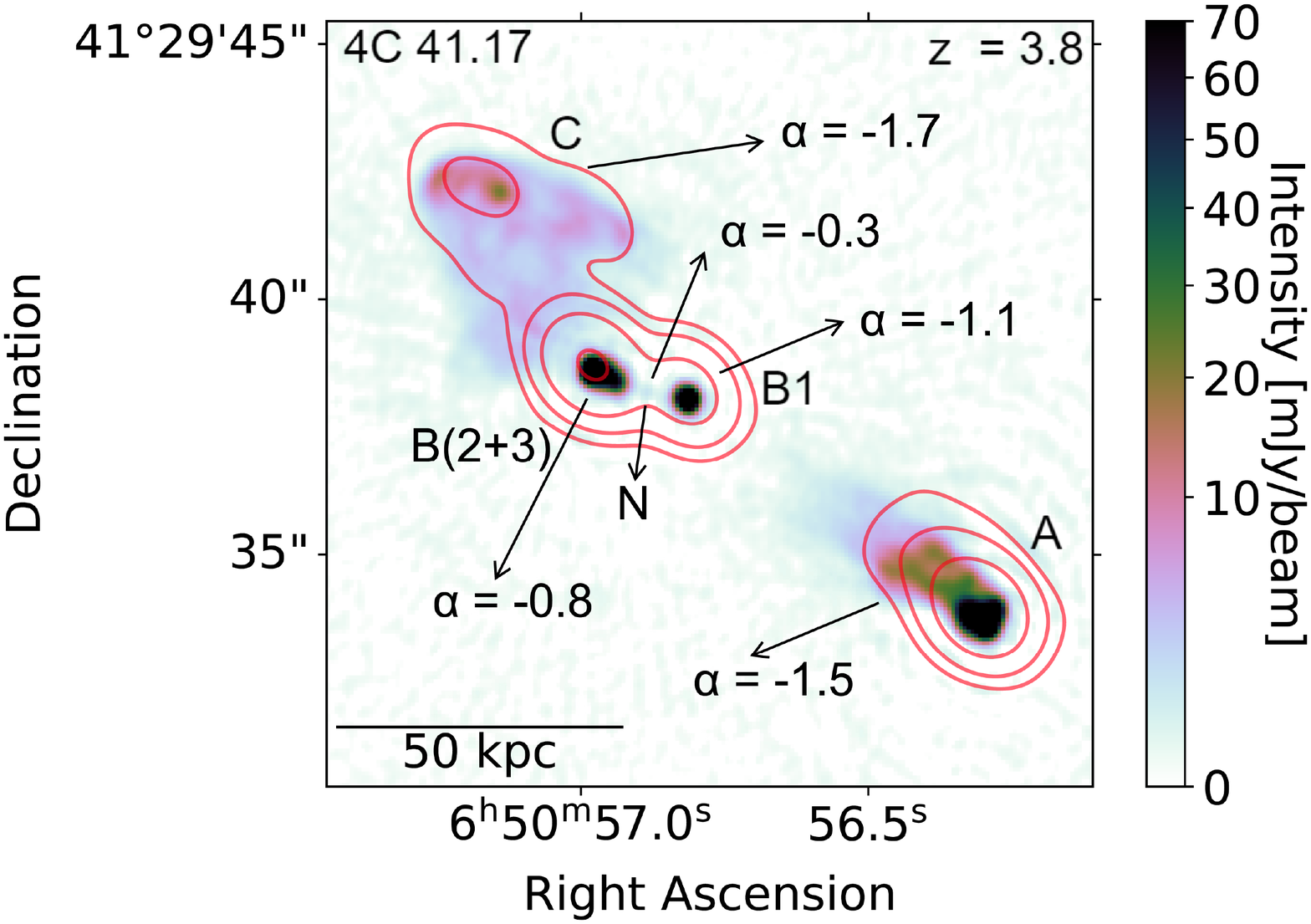}
  \caption{The ILT HBA 144~MHz final processed image of the Anthill at \(z=3.8\), with a resolution of 0.2" and the VLA contours in L band.}
     \label{fig:LOFAR_briggs_m1_4C4117.pdf}
\end{subfigure}
\begin{subfigure}
 \centering
\includegraphics[width=0.9\linewidth]{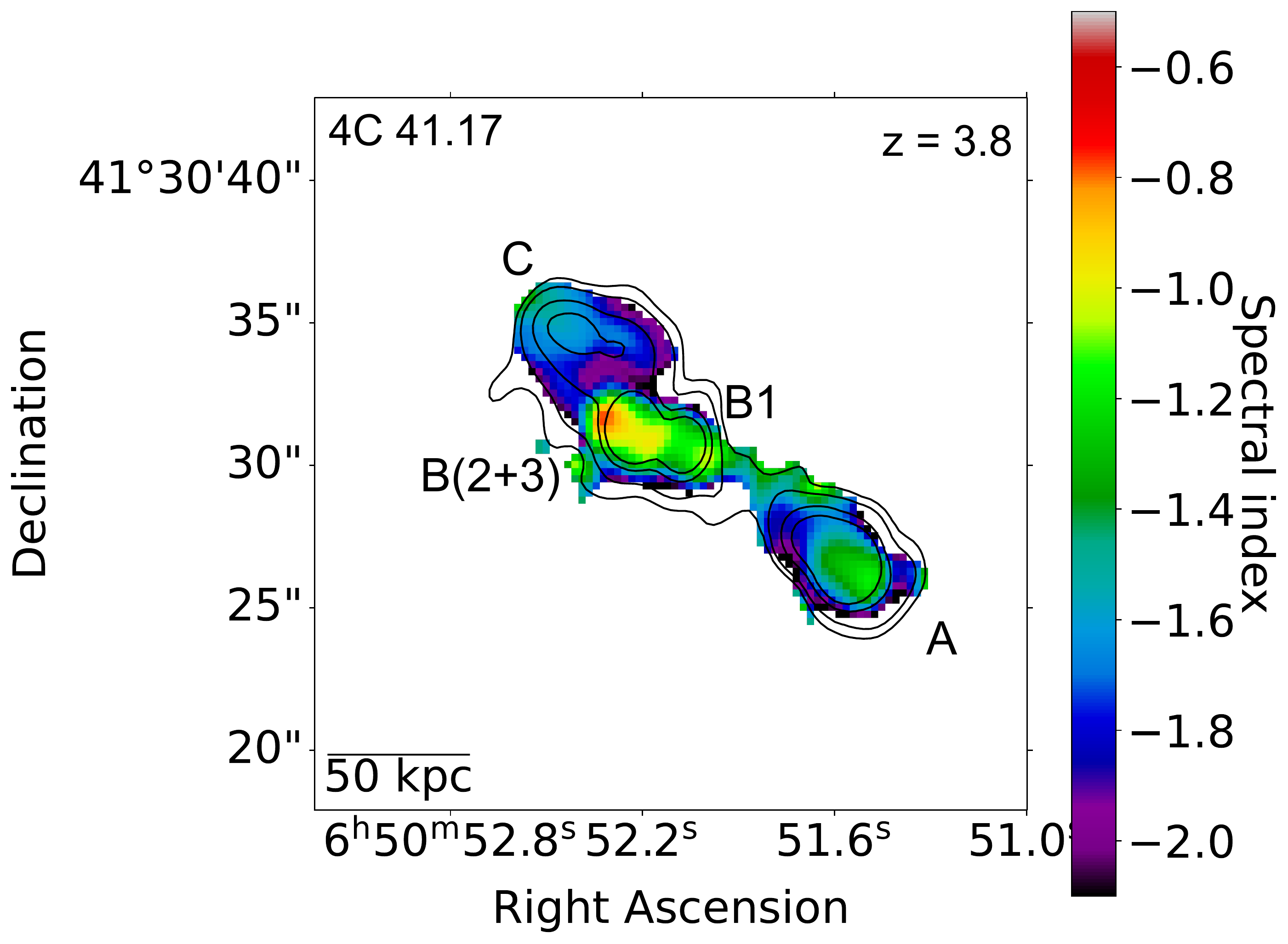}
\caption{Map of the low frequency 144~MHz - 1.4~GHz spectral index of the Anthill at \(z=3.8\) obtained using the ILT HBA and VLA L-band observations. The ILT HBA contours from the smoothed 144~MHz image in Figure~\ref{fig:LOFAR_briggs_m1_4C4117.pdf} are also displayed on the spectral index map.}
\label{fig:SI 4C41.17}

\end{subfigure}
\end{figure}

\begin{figure}
\begin{subfigure}
    \centering
    \includegraphics[width=0.9\linewidth]{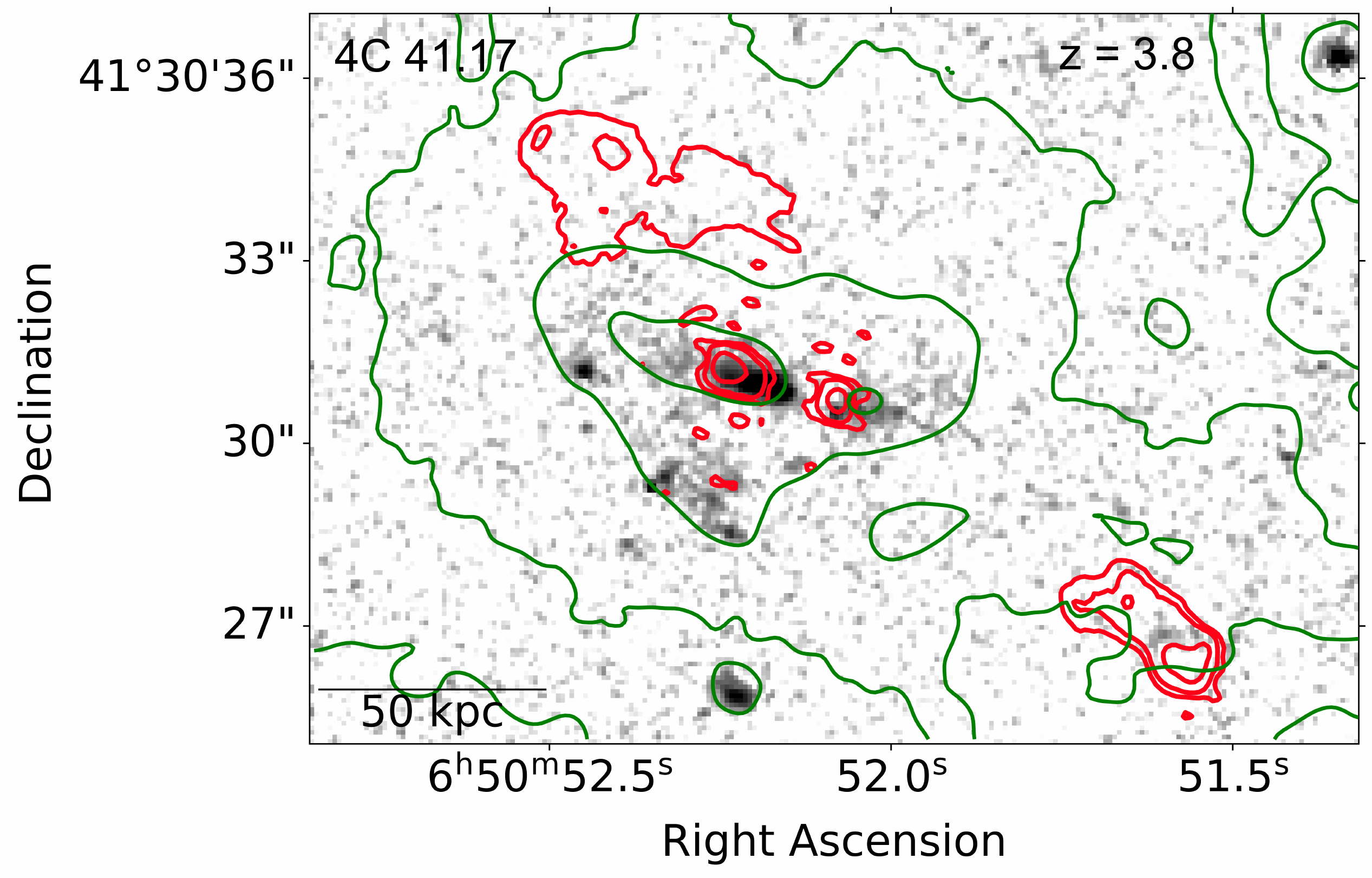}
\end{subfigure}
\begin{subfigure}
  \centering
    \includegraphics[width=0.9\linewidth]{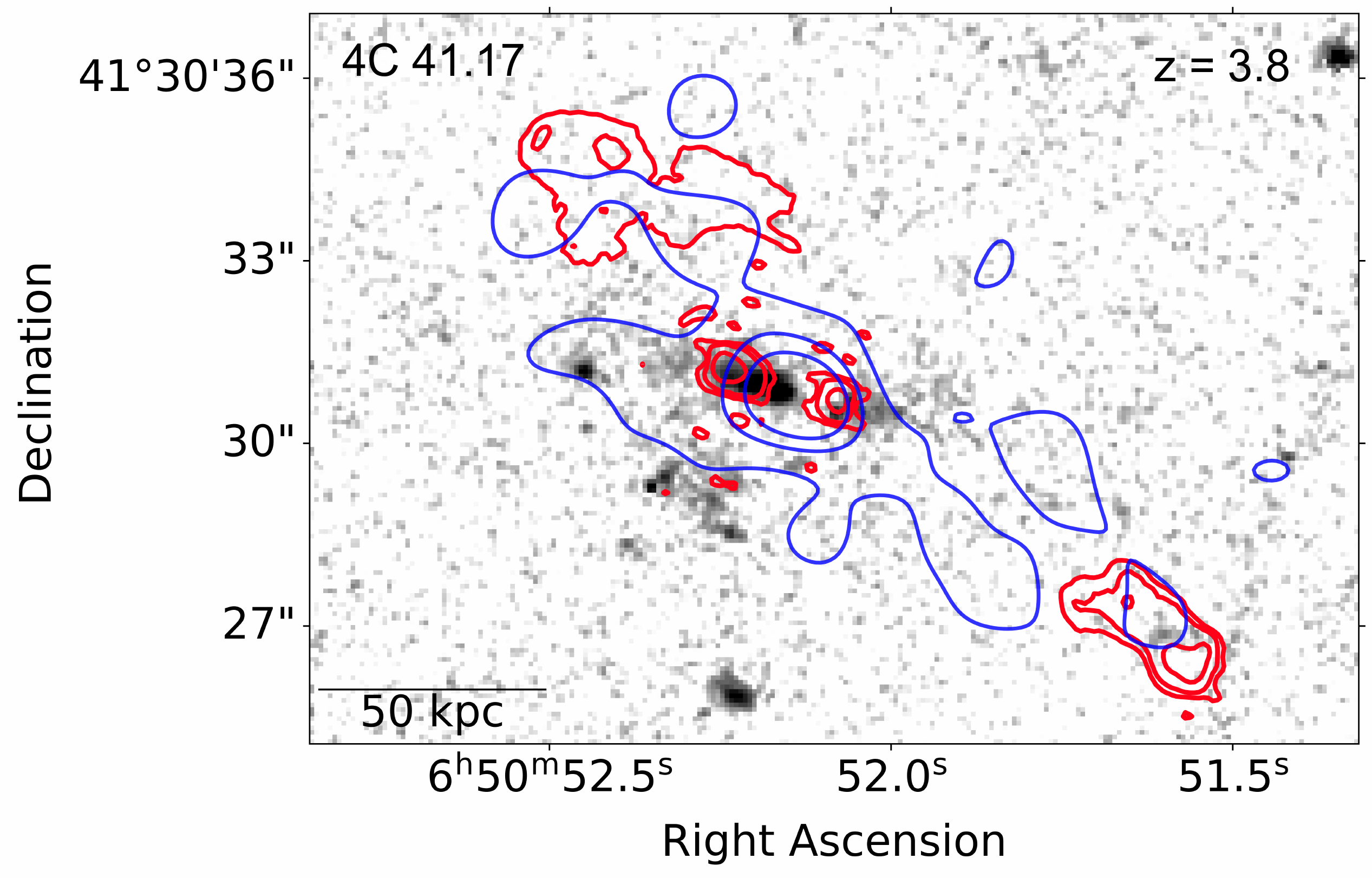}
\end{subfigure}

\caption{Overlay of ILT 144~MHz radio image for 4C\,41.17 at \(z=3.8\) (red contours) above 3 $\sigma$ noise, superimposed on the HST continuum image \citep[4~hour exposure through the F702W filter of WFPC2][]{bicknell2000jet}. The green contours (top panel) add the Keck Ly$\alpha$ distribution observed by Keck \citep{Reuland_2003}, and the blue contours (bottom panel) represent the X-ray distribution observed by Chandra \citep{scharf2003extended}.}
\label{fig:overlay4C41.17}
\end{figure}

The high-resolution image of the Anthill at 144~MHz is shown in Figure~\ref{fig:LOFAR_briggs_m1_4C4117.pdf} and its low-frequency spectral index map in Figure\ref{fig:SI 4C41.17}. Figure~\ref{fig:overlay4C41.17} shows an overlay of the ILT radio map on the HST optical continuum image \citep{miley1992hubble} together with Keck Ly\(\alpha\) contours (green, top) and Chandra X-ray contours (blue), bottom.

The figures illustrate that the Anthill has several interesting features. First note that the individual radio components designated at higher frequencies as A, B1, and C by \citet{chambers19904c}, as well as the nucleus (N) detected by \citet{carilli1994radio}, are distinguishable in our 144~MHz ILT map. However, B2 and B3 are blended at the ILT resolution and we shall refer to them as B(2+3). The radio source has an edge-brightened FR-II morphology similar to Cygnus\ A, with a pair of hot spots at its eastern edge (component C) and a brighter hot spot with a symmetric double narrow feature close to its western edge (Component A). There are double features on both sides of the source. This is apparent in the outer pair of eastern hot spots and the bifurcated backward jet-like structure in the western hot spot. This may be coincidental or could be a consequence of their production in the same episode of enhanced nuclear activity. The two pairs of jets could be produced in different activity episodes, with the jet angle between them corresponding to the precession of the radio source \citep{blundell1999inevitable}. However, we suggest that the complex shape of the source is more likely due to interaction between the radio jets and the gaseous halo around the host galaxy, such as is commonly observed to occur in HzRGs \citep{bicknell2000jet}.

Although the integrated low-frequency spectral index of the Anthill is ultra-steep with $\rm \alpha=-1.2$, there are substantial differences in spectra between the components. The outer components are steepest, with C having $\rm \alpha = -1.60 \pm 0.04$ and A having $\rm \alpha = -1.40 \pm 0.04$. The central compact component (B1) has a spectral index of $\alpha=-1.10 \pm 0.04$, while the offset component, B(2+3), features a flatter spectral index of $\alpha=-0.80 \pm 0.04$. Finally, we detect a faint 1.2~mJy brightness peak at the position of component N, which was previously detected by \citet{chambers19904c} and \citet{carilli1994radio}. This implies a spectral index of $\alpha\sim-0.3$ between our 144~MHz measurement and the flux density at 4.7~GHz as reported by \citet{carilli1994radio}.

Components B1 and B(2+3) are associated with a complex of objects in the optical image that are consistent with merging galaxies in the process of building a massive CD galaxy at the center of the cluster. Their optical emission is likely to be due to some combination of AGN and star formation processes.

An intriguing question is the nature of component B(2+3). Although only slightly resolved on the ILT image, it is resolved into two components on the VLA map.  B(2+3) is not aligned with the main collimated radio source structure (A, B1, C) and is co-located with an optical galaxy close to the massive central one. We note that the Anthill is a radio galaxy in which the forming central massive protogalaxy appears to be growing due to merging with several surrounding galaxies. B(2+3) may therefore well be associated with a second radio galaxy in the cluster that is merging with the central cD protogalaxy responsible for producing the main aligned (A, B1, C) radio source \citep{gurvits1997compact}. If the galaxy associated with B(2+3) is indeed merging with the central A, B1, C galaxy the fact that the radio jets (and SMBHs) in both galaxies are oriented in the same direction would be yet another observed alignment in orientation that could have fundamental implications for the origin and growth of SMBHs.

As described by \citet{miley2008distant}, the HST optical image shows that a group of several galaxies is associated with the 4C\,41.17 radio source, indicating that the Anthill is at the center of a rich galaxy cluster which is in the process of building a massive cD galaxy at its center. The whole complex is surrounded by a giant extended $Ly\alpha$ halo and an extended region of X-ray emission. The radio source, cluster galaxies, Ly$\alpha$ emission, and X-ray emission are all approximately aligned with each other, similar to in the Spiderweb radio protocluster \citep{carilli2022x}. The alignments indicate that there is ongoing interaction between the formation of the galaxy group and possible cluster and that the direction of the jets produced by the supermassive black holes associated with the jets is in some way connected with the large-scale formation processes. In particular, the alignment between radio jets and optical components, and the complex shape of the source, are indications of strong interactions between the jets and the surrounding gaseous halo, which is causing jet-induced star formation \citep{dey1997triggered,bicknell2000jet}.

The correspondence of the low-frequency radio emission and the X-ray images \citep{scharf2003extended} is consistent with the X-ray emission being dominated by inverse Compton up-scattering of the cosmic microwave background photons by the radio-emitting relativistic electrons. This is also similar to the case of the Spiderweb radio protocluster \citep{carilli2022x}.


\subsection{B2\,0902+34 at \(z=3.4\)}
\begin{figure}[H]
\centering
\begin{subfigure}
  \centering
  \includegraphics[width=0.86\linewidth]{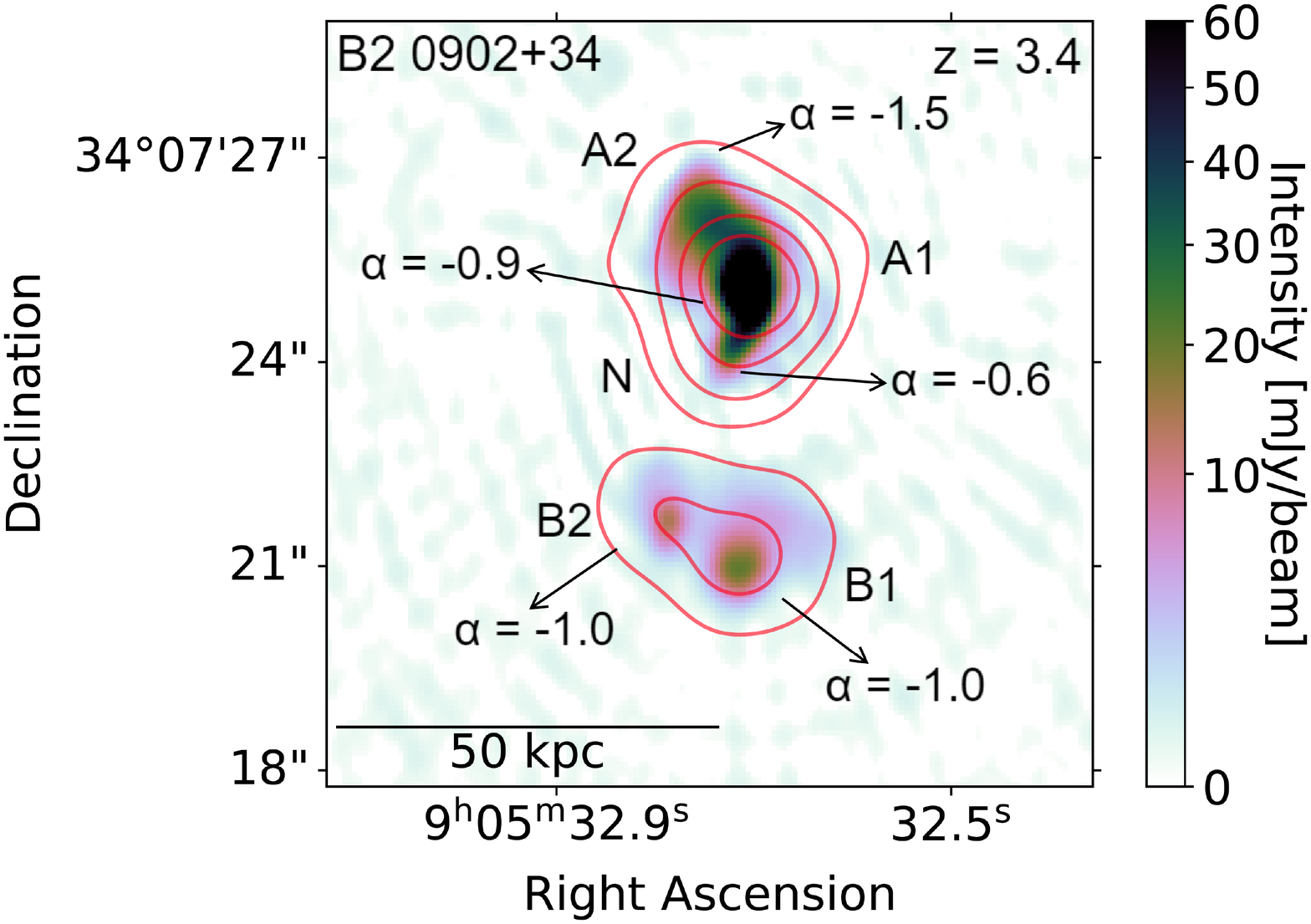}
  \caption{The ILT 144~MHz final processed image of B2\,0902+34 at \(z=3.4\), with a resolution of $0.4" \times 0.2"$ and the VLA contours in L band.}
     \label{fig:LOFAR_B0902}
\end{subfigure}
\begin{subfigure}
  \centering
  \includegraphics[width=0.88\linewidth]{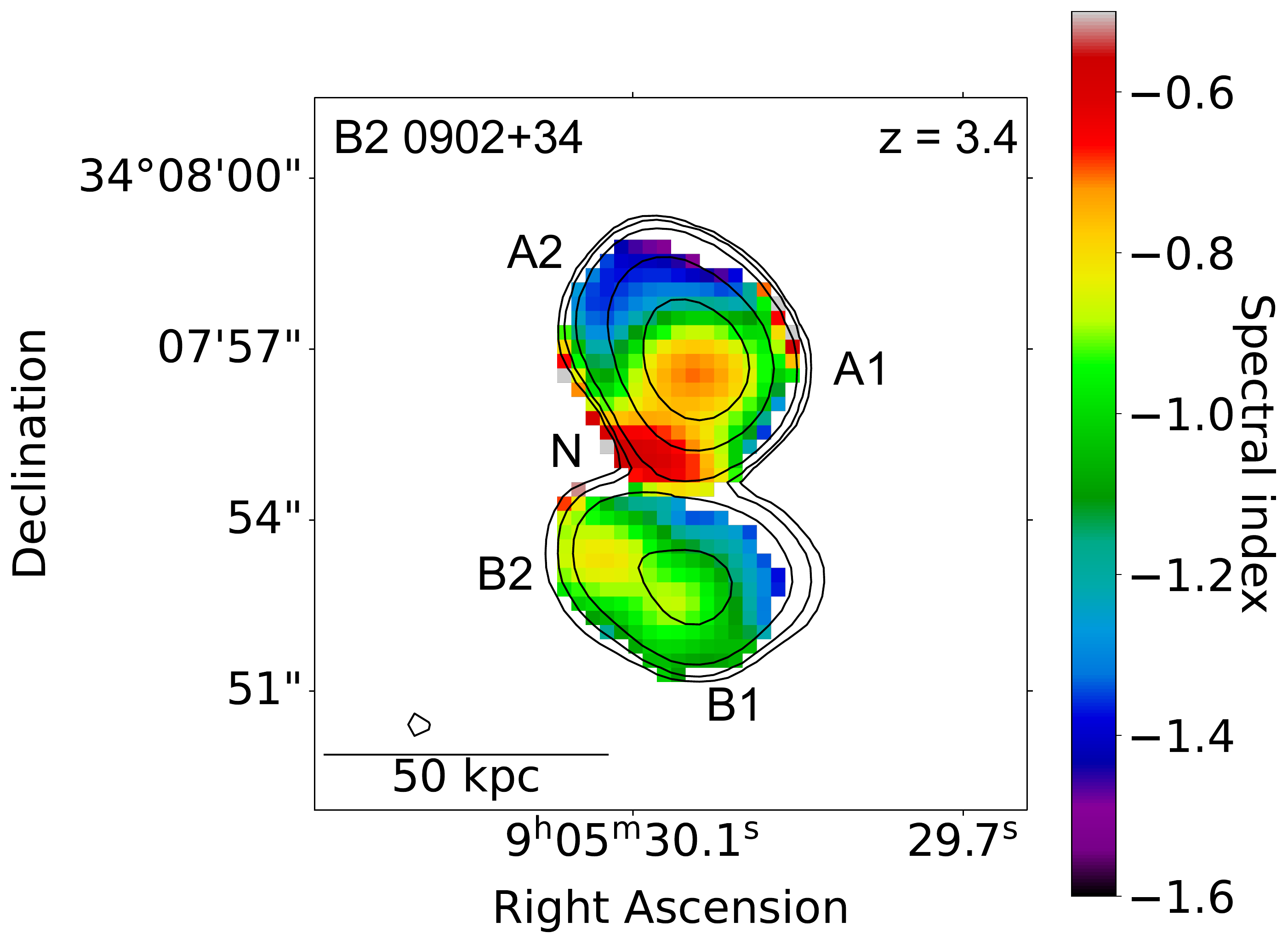}
  \caption{Map of the low frequency 144~MHz - 1.4~GHz spectral index of B2\,0902+34 at \(z=3.4\), obtained using the ILT HBA and VLA L-band observations. The ILT HBA contours from the smoothed image in Figure~\ref{fig:LOFAR_B0902} are displayed on the spectral index map.}
\label{fig:SI B20902}
\end{subfigure}
\begin{subfigure}
  \centering
  \includegraphics[width=0.78\linewidth]{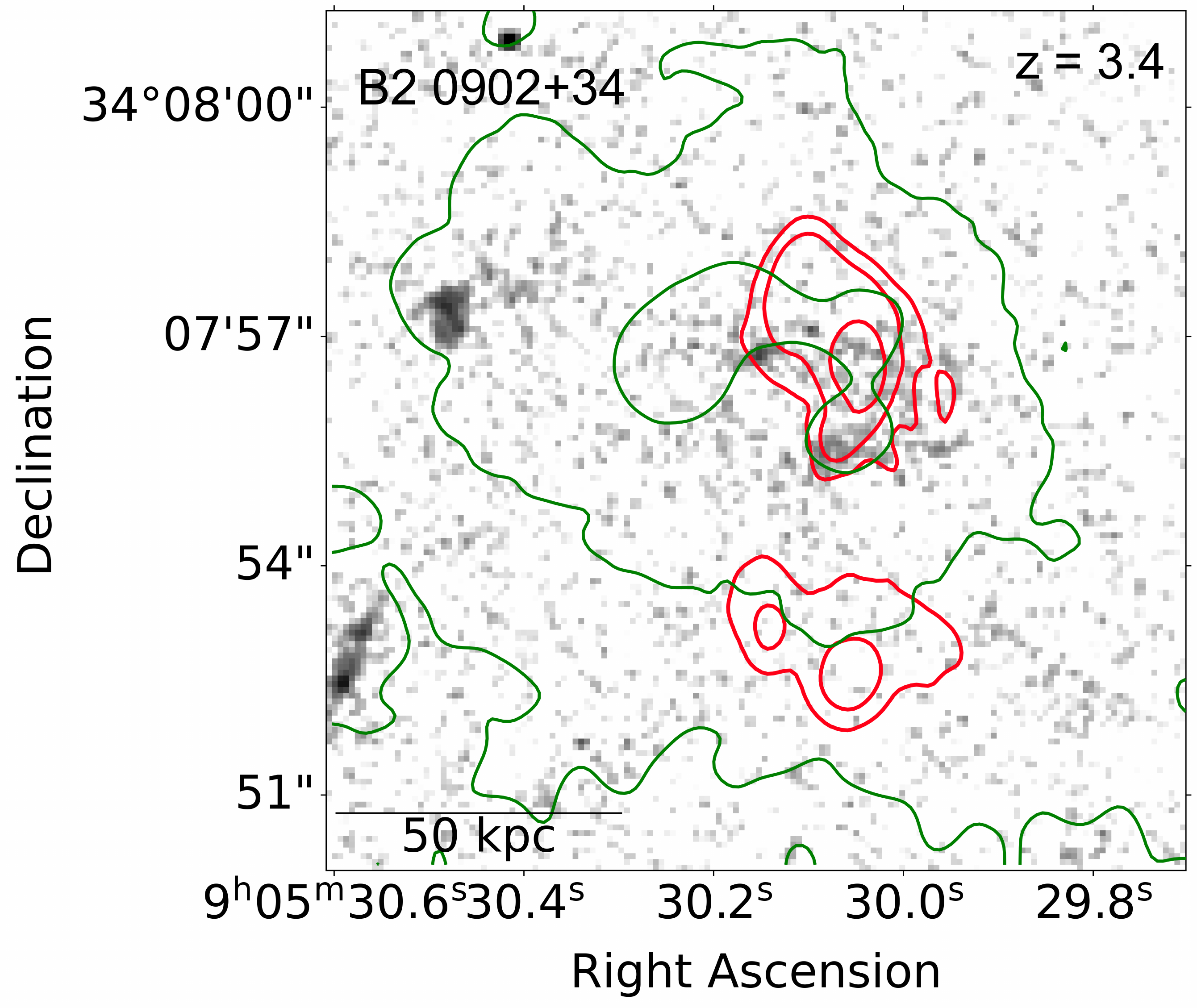}
  \caption{Overlay of LOFAR ILT radio image for B2\,0902+34 at \(z=3.4\) (red contours) superimposed on the HST continuum image \citep[6~hour exposure, PC section of WFPC2, F622W filter,][]{pentericci1999hst}. The green contours show the Keck Ly$\alpha$ distribution \citep{Reuland_2003}}
\label{fig:overlay_B20902}
\end{subfigure}
\end{figure}

The high-resolution image of B2\,0902+34 at 144~MHz is shown in Figure~\ref{fig:LOFAR_B0902} and its low-frequency spectral index map in Figure~\ref{fig:SI B20902}.  Figure\ref{fig:overlay_B20902} shows an overlay of the ILT radio map on the HST optical continuum image \citep{pentericci1999hst} together with the Ly$\alpha$ contours as observed by the Keck Telescope \citep{Reuland_2003}.

There are several aspects of the low-frequency radio image and spectral index map of B2\,0904+32 that deserve comment. The low-frequency ILT image consists of multiple components: the nuclear core N, the northern component A1 and A2, with an ultra-steep spectrum radio plume having an ultra-steep spectrum with $\alpha=-1.5$ at its northern edge and a southern lobe with two outer hot spots \citep[B1 and B2; the naming is inspired by][]{carilli1995bizarre}). The nuclear core has a spectral index between 1.4~GHz and 144~MHz of $\alpha=-0.60 \pm 0.02$. Although relatively flat, it is steeper than $\alpha=-0.30 \pm 0.02$ measured by \citet{carilli1995bizarre} between 1.65~GHz and 4.70~GHz. It, therefore, appears to have a peaked  \textquotedblleft GPS-type\textquotedblright spectrum. There is a pair of similar lobes (components B1 and B2) and a highly bright lobe (A1) that extends to the east. Just like for 4C 41.17, the lobes could have appeared from different activity episodes, where the precession of the radio source  is jet angle between them \citep{blundell1999inevitable}. However, a more likely explanation for the complex shape of the source is the interaction of the radio jets with the surrounding halo, similar to 4C~41.17.

B2\,0902+34 is located in a protocluster environment \citep{eales1993evidence, pentericci1999hst}, surrounded by several galaxies. As pointed out by \citet{carilli1995bizarre}, B2\,0902+34 has an unusual radio structure, with a relatively flat-spectrum core, located in a  \textquotedblleft valley\textquotedblright between 2 optical peaks.

The bending of the northern structure eastwards and the southern structure westwards indicates that the jet ejected from the nucleus in the A1 - B2 direction is thereafter bent because of its interaction with a rotating gas cloud connected with the Ly\(\alpha\) halo visible on the Keck infrared map Ly$\alpha$ map of \citep{Reuland_2003} in Figure \ref{fig:overlay_B20902}. The ultra-steep spectrum part of the source (A2) coincides with a region shown by \citet{carilli1995bizarre} to have large rotation measures and gradients in rotation measure that are indicative of a dense, magnetized gas cloud, comparable to those found at the centers of cooling-flow clusters at low redshifts.

\subsection{4C\,34.34 at \(z=2.4\)}
\begin{figure}[ht]
\centering
\begin{subfigure}
  \centering
  \includegraphics[width=0.88\linewidth]{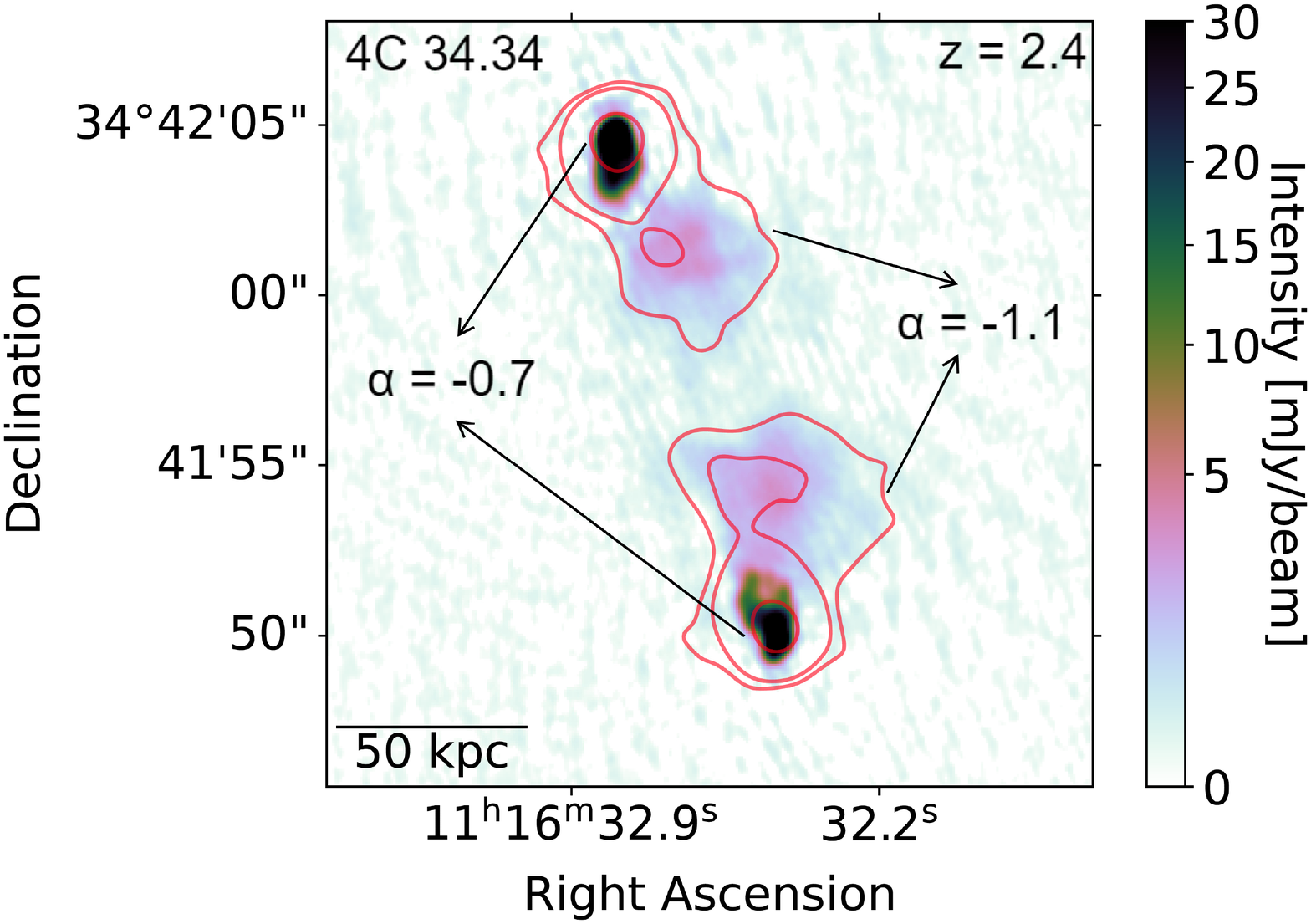}
  \caption{The ILT 144~MHz final processed image of 4C\,34.34 at \(z=2.4\), with a resolution of $0.3" \times 0.2"$ and the VLA contours in L band.}
     \label{fig:LOFAR_4C3434}
\end{subfigure}
\begin{subfigure}
  \centering
  \includegraphics[width=0.9\linewidth]{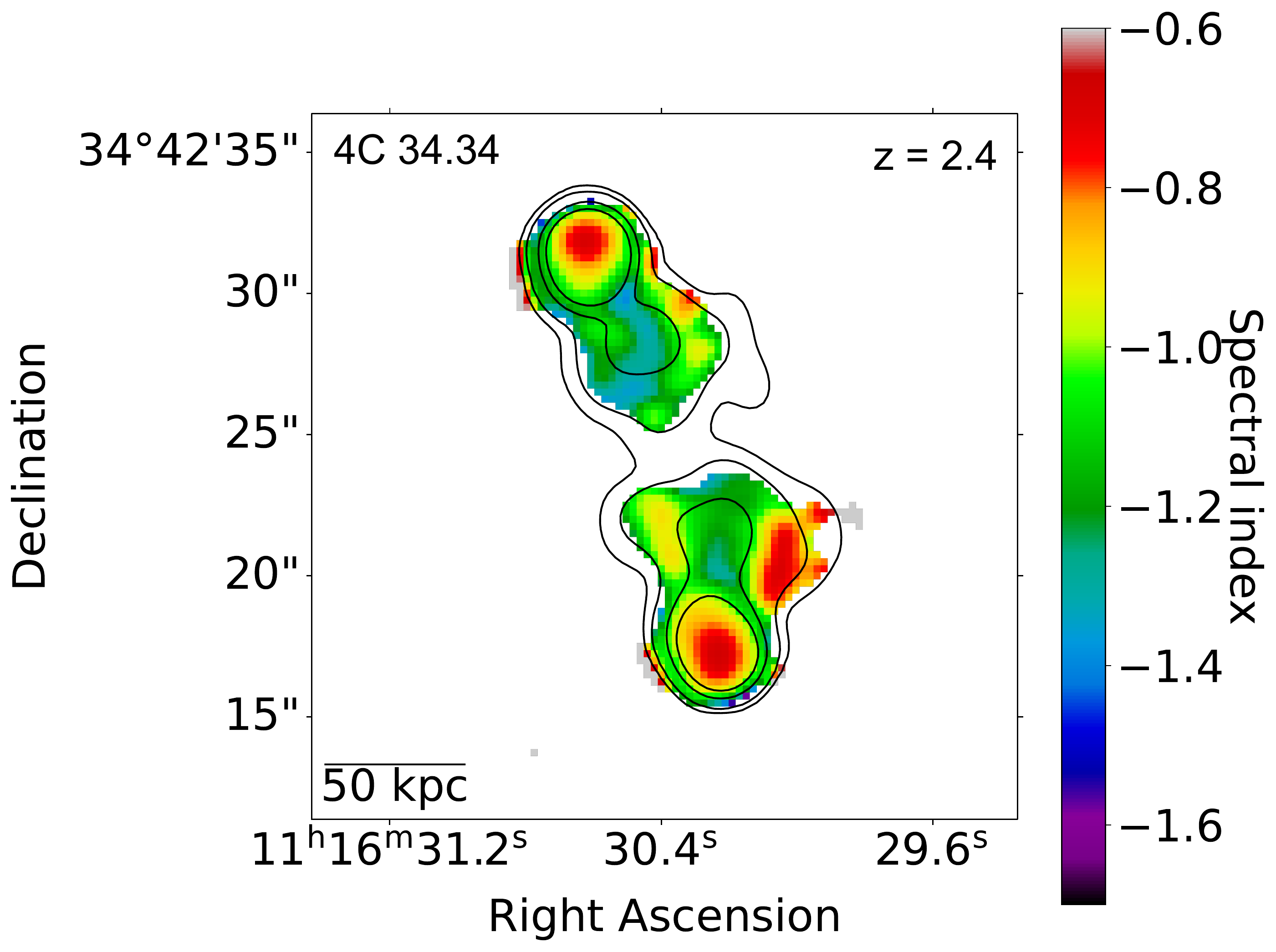}
  \caption{A map of spectral index of 4C\,34.34 at \(z=2.4\) obtained using VLA L-band and the ILT HBA observations. The ILT HBA contours from the smoothed image in Figure~\ref{fig:LOFAR_4C3434} are displayed on top of the spectral index map.}
\label{fig:SI 4C3434}
\end{subfigure}

\end{figure}

The high-resolution image of 4C\,34.34 at 144~MHz is shown in Figure~\ref{fig:LOFAR_4C3434} and the low-frequency spectral index map in Figure~\ref{fig:SI 4C3434}. It is not possible to make optical or X-ray overlays similar to those shown for the Anthill, Figure~\ref{fig:overlay4C41.17} and B2\,0902+34, Figure~\ref{fig:overlay_B20902} because no comparable non-radio data are available.

We note the following aspects of the low-frequency radio image and spectral index map of 4C\,34.34 at \(z=2.4\). The radio structure of 4C\,34.34 is typical of high-luminosity FR-II edge-brightened double radio sources, with hot spots at its northern and southern extremities and diffuse emission on both sides extending towards the galaxy nucleus that is not detected by the ILT. The spectral indices of the hot spots are normal for extended radio sources, both with $\alpha = -0.70 \pm 0.04$. The radio \textquotedblleft bridge\textquotedblright has an ultra-steep spectrum with $\alpha = -1.10 \pm 0.05$.

\subsection{4C\,43.15 at \(z=2.4\)}

\begin{figure}[!htb]
\centering
\begin{subfigure}
  \centering
  \includegraphics[width=0.88\linewidth]{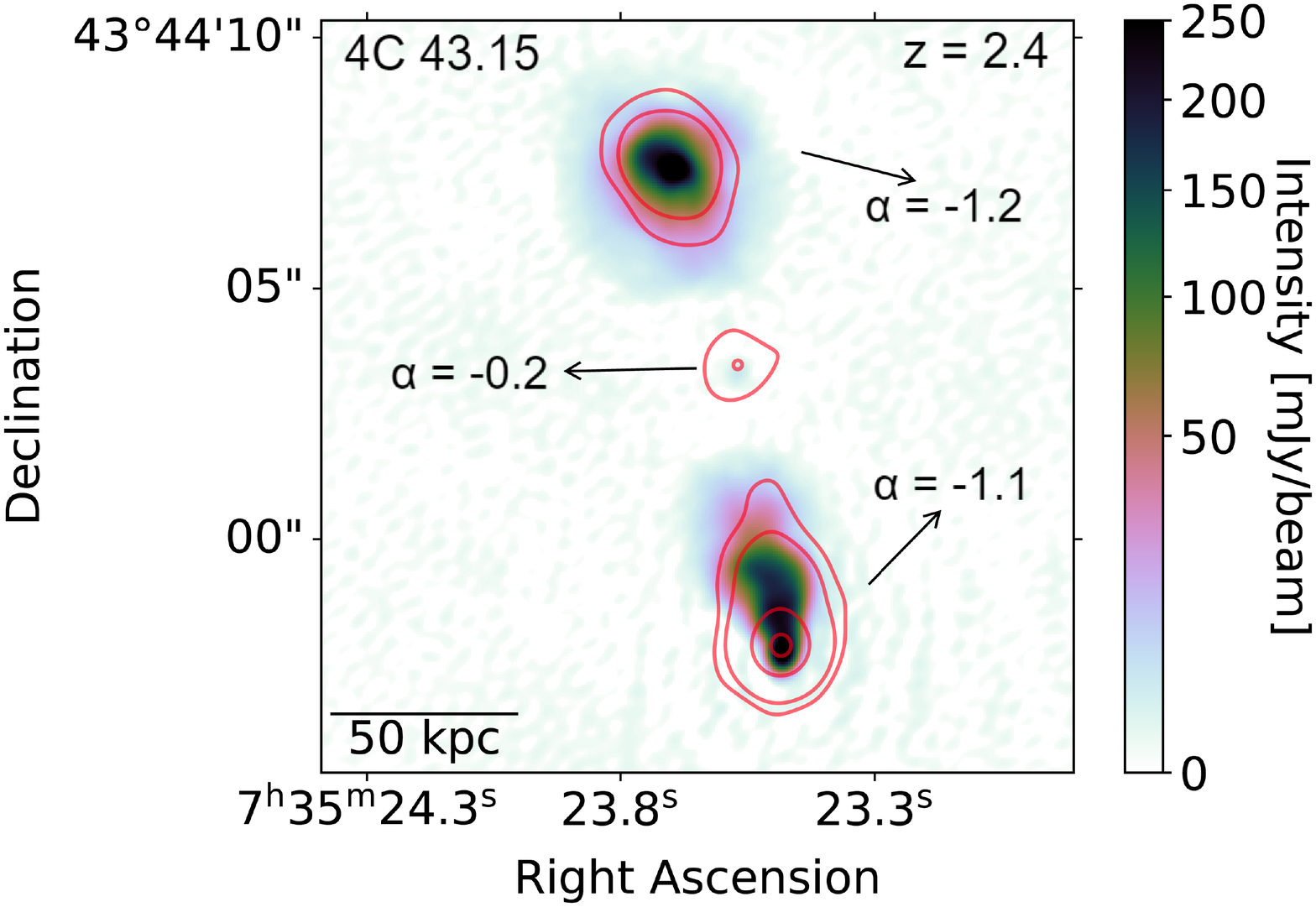}
     \caption{The ILT 144~MHz final processed image of 4C\,43.15 at \(z=2.4\), with a resolution of $ 4" \times  4"$ and the VLA contours in X band.}
     \label{fig:LOFAR_4C4315}
\end{subfigure}
\begin{subfigure}
  \centering
     \includegraphics[width=0.9\linewidth]{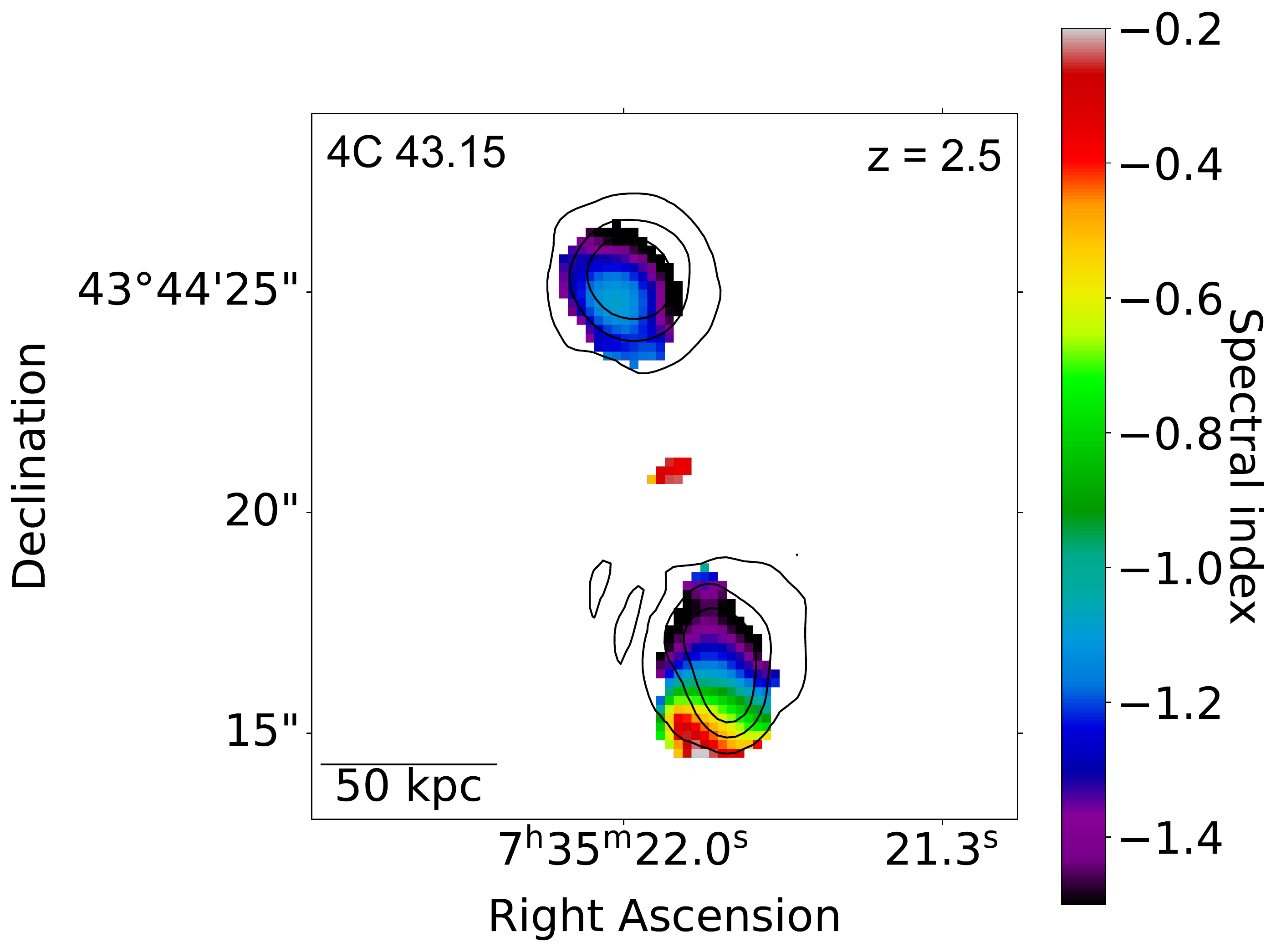}
\caption{A map of spectral index of 4C\,43.15 at \(z=2.4\) obtained using VLA X~band and the ILT HBA observations. The ILT HBA contours from the smoothed image from Figure~\ref{fig:LOFAR_4C4315} are displayed on top of the spectral index map.}
\label{fig:SI 4C4315}
\end{subfigure}

\end{figure}

The high-resolution image of 4C\,43.15 at 144~MHz is shown in Figure~\ref{fig:LOFAR_4C4315} and its low-frequency spectral index map in Figure~\ref{fig:SI 4C4315}. Both images were taken from \citet{sweijen2022high}. Again, no suitable data are available for making optical or X-ray overlays similar to those shown for the Anthill or B2\,0902+34.

We note the following aspects of the low-frequency radio image and spectral index map of 4C\,43.15 at \(z=2.4\). As with 4C\,34.34, the radio structure of 4C\,43.15 is typical for a high-luminosity FR-II edge-brightened double radio source, with hot spots at the northern and southern extremities. The spectral indices of the hot spots are normal, both with $\alpha = -0.70 \pm 0.02$ and that of the lobe is steeper, with $\alpha = -1.10 \pm 0.03$. There is a flat--spectrum radio core, bright on the VLA X-band image and faint on the ILT HBA image that has a spectral index of $\alpha = -0.40 \pm 0.07$. The similarity between 4C 34.34 and 4C 43.15 is high, both in their radio images and their spectral indices.

A Ly$\alpha$ image \citep{motohara2000infrared} and the Ly$\alpha$, NV, CIV, and HeII spectra \citep{villar2003kinematically} show that an ionized gas halo extends asymmetrically to the north and envelopes the northern radio hot spots. The southern hot spot is located outside the ionized gas halo and is more compact than the northern one. This may be because the gas is interacting with and compressing the northern jet during its outward propagation.

\section{Discussion}
The sub-arcsecond resolution LOFAR images of the four HzRGs discussed above illustrate the diversity of properties exhibited by luminous $z>2$ radio galaxies and the importance of their study in understanding the role and influence of AGN on the formation and evolution of massive galaxies and galaxy clusters.
\subsection{Radio structures as signatures of protoclusters} 4C\,41.17 (the Anthill) at \(z=3.8\) and B2\,0902+34 are both associated with radio structures that appear  \textquotedblleft bent\textquotedblright or non-linear and are identified with galaxies in dense groups or protoclusters. They are surrounded by giant halos of ionized gas. Their radio structures are non-typical for low-redshift high-luminosity radio sources and have probably been \textquotedblleft distorted\textquotedblright by the interaction of their radio jets  with the dense protocluster environment at high redshifts. A distorted radio structure has been suggested as a tool for finding protoclusters around high-z quasars by \citet{1988Natur.333..319B}.

4C\,34.34 at \(z=2.4\) and 4C\,43.15 have FR-II edge-brightened structures that are typical for high-luminosity radio sources associated with galaxies and quasars at lower redshift. They do not appear to be located in such dense protocluster environments as 4C\,41.17 or B2\,0902+34, despite the extended halo of ionized gas that is located asymmetrically with respect to the extended 4C\,43.15 radio source.

\subsection{Interaction of radio jets with the environment}

Radio-optical alignments similar to that observed in the Anthill (Figure~\ref{fig:overlay4C41.17}) are seen in many other \(z>3\) HzRGs \citep{miley2008distant} have been used as evidence that the radio jets in HzRGs can often interact vigorously with the surrounding gas. Several mechanisms have been proposed to explain this interaction, including positive AGN feedback and jet/cloud deflection. In positive AGN feedback scenarios, the jets induce star formation in the gas as they propagate outwards \citep{carilli1997radio, pentericci2000vla}, either by compressing the cold molecular clouds \citep{silk2013unleashing}, or directly in the out-flowing gas \citep{ishibashi2012active}. Jet/cloud deflection is seen in B2 0902+34 (A2 component) and takes place when the jet hits a highly dense environment, interacts with it, and can be deviated \citep{de1991deflection, dal1999three}. All these mechanisms may be important because they explain the influence of the jets on the interstellar and intergalactic medium around such massive galaxies. Multispectral studies of such interactions from a larger sample of HzRGs are needed to constrain the details of the processes that are occurring

The Anthill at \(z=3.8\) is one of the most spectacular case study of an emerging dense cluster, similar to but more distant than the Spiderweb at \(z=2.2\). The alignments of the radio emission with the (i) galaxy distribution, (ii) the giant ionized gas halo and (iii) the X-ray emission indicate the intriguing connection between the various constituents. 
The additional alignment of the radio jet of the central galaxy (A,B1,C) with the radio jet associated with the possibly merging satellite galaxy B(2+3) is particularly interesting.  It is additional evidence that at redshifts \(z>2\) supermassive black holes, massive galaxies, and massive galaxy clusters are built up together in fundamental processes that involve galaxy and SMBH merging and star formation stimulated by the outward-moving radio-emitting jets. 

\subsection{Inverse Compton scattering of radio to X-rays} The correspondence of the location and alignment of the radio emission and the X-ray emission in the Anthill is consistent with the X-ray radiation being dominated by inverse Compton up-scattering of the cosmic microwave background photons by the radio-emitting relativistic electrons as also seems to be the case in the Spiderweb protocluster \citep{carilli2022x, schwartz2002chandra, carilli2002cluster, scharf2003extended}. We note that because of the dependence of the energy density of the CMB on $(1+z)^4$, this would be a factor of $\sim 5$ larger at the redshift of the Anthill than at the redshift of the Spiderweb. However, the ratio of Inverse Compton X-ray emission to synchrotron radio emission is dependent on many unknown factors and assumptions, such as magnetic field strengths and pressures within the radio jets and hot spots \citep[e.g.][]{carilli2022x, hodges2021proof}, high spatial resolution at low radio frequencies are important for investigating such processes. The discovery of a luminous radio galaxy with \(z=5.72\) \citep[150~MHz luminosity $\rm 10^{29.1}$~$\rm W\ Hz^{-1}$ and spectral index -1.4][]{saxena2018mnras} implies that IC dimming of radio emission is not substantial for redshifts \(z<6\).

\subsection{High-z environment and the $\alpha$ -- z relationship} For both the Anthill at z = 3.8 and B2 0902+34 at z =3.4, the location of the ultra-steepest part of the radio emission coincides with the Ly $\alpha$-emitting ionized gas halo. This suggests that the correlation of integrated ultra-steep spectral index with redshift discussed in the introduction may be at least partially due to the location of the synchrotron jets in the dense environment associated with forming protoclusters in the early Universe \citep{klamer2006search}. A statistical comparison of the spatial relationship of radio spectral index with the various resolved constituents of distant radio galaxies will be important in future ILT studies with larger samples. 

\section{Conclusions and future prospects}

This pilot project has shown that sub-arcsecond studies of the low-frequency radio structures of radio galaxies at \(z>2\) with the ILT are feasible and highly informative. Because LOFAR probes the oldest detectable radiating synchrotron electrons and because the extended radio emission contains information about the activity history of the active host nuclei and supermassive black holes, such studies can provide unique information about the formation and evolution of massive galaxies and protoclusters in the early Universe.

HzRGs have several properties that make high-resolution studies with LOFAR uniquely suited to searching for the most distant massive protoclusters and studying how the protocluster AGN and their synchrotron jets interact with the forming parent protoclusters. These include (i) ultra-steep radio spectra that allow LOFAR to detect the oldest detectable radiating synchrotron electrons, (ii) extended radio emission with an angular size distribution well suited to mapping with the international baselines of the ILT \citep[80\% between 4 and 20"; e.g.,][]{1998PhPl....5.1981C}, (iii) a structure that is frequently bent or non-linear due to interaction with the extended ionized gas halos that permeate distant radio protoclusters.

There are three fundamental diagnostics of HzRGs which require the high resolution at low frequencies that are attainable with the ILT.
\begin{itemize}
    \item Establishment that HzRGs have bent, nonlinear, and asymmetric radio structures can indicate that the associated radio sources are likely to be interacting with a dense protocluster environment. This can pinpoint the location of massive protoclusters and provide important information about their properties   \citep{1988Natur.333..319B,battye2009radio}. It is likely that linear, highly symmetrical sources are located in less dense environments, not specific to protoclusters.

\item The comparison  of spatially resolved low frequency radio images to high frequency radio images allows us to calculate the spatially resolved spectral index, and in some cases (e.g. 4C43.15) conduct spectral modeling and find out the spectral age of the radio emitting plasma.
    
 \item Comparison of the sub-arcsecond spatial distribution of the radio images and radio spectra with images at non-radio wavebands. These are needed to understand how the forming supermassive black holes and their jets are related to merging, star-forming and other processes involved in building the first massive galaxies and protoclusters.
 \end{itemize}

This pilot project illustrates the potential importance  of extreme USS radio sources from the LOFAR-WHT-WEAVE LoTSS surveys \citep{shimwell2019lofar} for advancing such studies. Filtered to match the steep spectral indices, angular size distributions, flux ranges, and morphologies of known HzRGs and radio protoclusters, LOFAR samples will enable dense protoclusters like the Spiderweb and the Anthill to be detected out to the highest redshifts at which they emerged.  For example, a radio source with a similar luminosity and spectral index to the Anthill could be detected at high resolution with the ILT to redshifts beyond $z\sim$ 8. Hence samples of USS radio sources from LoTSS form a unique basis for multispectral investigations of interactions between powerful radio jets and the other constituents of protoclusters. Followup multispectral studies are likely to provide fundamental information about the simultaneous evolution of massive galaxies, dense protoclusters and supermassive black holes in the early Universe. 

Sub-arcsecond observations of USS radio protoclusters with the ILT would benefit by even higher resolutions than are presently available and such resolutions are feasible. We, therefore, suggest that extending the ILT baselines by a substantial factor and the addition of ILT stations at strategic locations to achieve more uniform baseline coverage and increased sensitivity should be explored. 

\begin{acknowledgements}
We thank Chris Carilli for his useful comments.
RJvW and RT acknowledge support from the ERC Starting Grant ClusterWeb 804208. 

This paper is based (in part) on data obtained with the International LOFAR Telescope (ILT). LOFAR \citep{vanhaarlem13} is the Low Frequency Array designed and constructed by ASTRON. It has observing, data processing, and data storage facilities in several countries, that are owned by various parties (each with their own funding sources), and that are collectively operated by the ILT foundation under a joint scientific policy. The ILT resources have benefitted from the following recent major funding sources: CNRS-INSU, Observatoire de Paris and Universit\'e d'Orl\'eans, France; BMBF, MIWF-NRW, MPG, Germany; Science Foundation Ireland (SFI), Department of Business, Enterprise and Innovation (DBEI), Ireland; NWO, The Netherlands; The Science and Technology Facilities Council, UK; Ministry of Science and Higher Education, Poland. The National Radio Astronomy Observatory is a facility of the National Science Foundation operated under cooperative agreement by Associated Universities, Inc. The J\"ulich LOFAR Long Term Archive and the German LOFAR network are both coordinated and operated by the Jülich Supercomputing Centre (JSC), and computing resources on the supercomputer JUWELS at JSC were provided by the Gauss Centre for Supercomputing e.V. (grant CHTB00) through the John von Neumann Institute for Computing (NIC).

\end{acknowledgements}

%
%

\bibliographystyle{aa}
\bibliography{bibliography}

\end{document}